**Radiation Shielding Performance of Different Concrete Materials: A Systematic Review**

[1*]Christiana Subaar, [2]Ziem Samuel Aanoneda, [3]Sylivia Boateng, [4]Emmanuella Konadu Amaniampong, [5]Philimon Adjei

[1,2,3,4,5] Kwame Nkrumah University of Science and Technology, College of Science, Department of Physics, Kumasi, Ghana.

**Corresponding author:** *Christiana Subaar
Kwame Nkrumah University of Science and Technology,
College of Science, Department of Physics, Kumasi, Ghana.
Phone Number: +233 207911602
Email: christiana.subaar@knust.edu.gh; ysubaar@gmail.com

**Abstract**
**Background:**
Concrete is one of the most-used material today in nuclear, medical, and industrial applications for radiation shielding due to its economic advantages and availability together with its structural performance. However, differences in the use of aggregates, density, and other additives affect radiation attenuation efficiency. It is therefore necessary to understand and compare shielding properties of various concrete formulations for the optimization of safety and performance in radiation-prone environments.
**Methods:**
Using Preferred Reporting Items for Systematic reviews and Meta-Analyses (PRISMA) guidelines, a literature search was performed across PubMed, Scopus, ScienceDirect, and Google Scholar. 17 peer-reviewed studies published between 2010 and 2025 were analysed systematically. Data was extracted based on material composition, density, radiation type, energy range, attenuation coefficients, and shielding efficiency. The obtained results were compared to find the trend in performance and optimization of the considered materials.
**Conclusion:**
Radiation shielding efficiency of concrete is dependent on its density, microstructural characteristics and type of aggregate. For superior performance in a mixed radiation field, Heavy and boron-rich additives can be added. Newly developed UHPCs and nano-engineered concretes are lightweight, durable, and environmentally friendly options for shielding materials compared to traditional ones. Further studies are needed to focus on the standardization of test methods, validation of long-term stability, and coupling computational modelling with experimental data in order to guide material design for applications featuring enhanced radiation shielding.

# 1.0 Introduction

The ever-increasing applications of ionizing radiation in medicine, industry, and energy require efficient shielding with respect to safety concerns regarding human beings and environmental conditions (Wu and Wang, 2024). Ionizing radiation involves gamma rays, X-rays, and neutron radiation that have enough energy to ionize atoms and molecules, therefore poses great health and structural risks (Apte and Bhide, 2024). The objective of radiation shielding is the attenuation of this sort of radiation to acceptable limits that ensure safety in nuclear power plants, radiotherapy centers, research laboratories, and space missions (Shultis and Faw, 2010).

Concrete has become one of the most common materials used for radiation shielding, since it is available, mechanically strong, and economical (Kharita et al., 2008). Unlike metals like lead or steel that are very expensive and cumbersome to handle, concrete offers design flexibility and structural functionality with similar efficiency as a radiation barrier. Its composition can be adjusted by changing the types and proportions of cements, aggregates, and additives to suit particular shielding requirements. As such, concrete is very important in big projects such as reactor containment buildings, diagnostic imaging bunkers, and storage facilities for radioactive waste.

The effectiveness of concrete for the purpose of radiation shielding is influenced by material composition, density, and thickness (Jóźwiak-Niedźwiedzka and Lessing, 2019). This paper reviews specific research that used high-density aggregates like barite ($BaSO_4$), magnetite ($Fe_3O_4$), hematite ($Fe_2O_3$), and ilmenite ($FeTiO_3$), which increases the mass attenuation coefficient and linear attenuation coefficient ($\mu$), giving a significant rise in attenuation capability (Ahmad et al., 2019). Heavyweight concretes with densities between 3,000 to 5,000 kg/m³ show better performances compared to normal-weight concrete with a density of 2,300–2,400 kg/m³ (Özen et al., 2016). Development in concrete technology has introduced materials like ultra-high-performance concrete with superior mechanical strength and low porosity, due to which their shielding performance is enhanced (Abdal et al., 2023). The dense microstructure of UHPC minimizes the voids and microcracks within the structure. This reduces the photon and neutron penetration within the structure, hence providing improved durability under radiation.

The shielding capability of the concrete against radiation is generally expressed in terms of linear attenuation coefficient (μ), half-value layer (HVL), and tenth-value layer (TVL) (Ahmad et al., 2019). The linear attenuation coefficient expresses a material's ability to absorb or scatter radiation, the magnitude of which would increase with an increase in the material's shielding capability (Halliwell, 2021). HVL and TVL represent the thickness at which a material could reduce the intensity of radiation to 50% and 90% respectively (Abouelnour et al., 2025). All these properties are functions of some parameters, including density, atomic number Z of constituent elements, and energy of incident radiation. Theoretically, materials with higher atomic numbers and densities provide better protection against gamma and X-rays, while for neutron attenuation, hydrogenous or boron-based compounds must be added to the matrix (Duru et al., 2024).

Recent research efforts have been directed toward optimizing concrete mixture as a means to heighten shielding efficiency with no compromise to structural performance (Abdullah et al., 2022). Substitution of conventional aggregates with heavy minerals such as barite, magnetite, or ilmenite results in increasing density, while at the same time reducing the HVL value (Tesaye and Dinku, 2023). As an example, using barite concrete can achieve up to 30-40% higher attenuation than ordinary concrete in the middle- to high-energy gamma rays. Radiation-proof UHPC is prepared by modifying UHPC; it involves reinforcement with heavy mineral or nano-additives like bismuth oxide, which significantly improves its radiation resistance without affecting its high compressive strength and durability. These studies represent a big step toward multifunctional concretes that combine structural robustness with superior shielding capabilities.

Apart from material design, the effectiveness of shielding is also affected by the nature of radiation, its intensity, and duration of exposure (Shazad et al., 2022). While high-energy gamma radiation calls for high-density materials containing elements of high atomic number Z, neutron shielding is favoured by hydrogen-rich or boron-containing compounds (Abdulrahman et al.,2020). Sustainability issues have recently fostered interest in the use of industrial by-products such as fly ash, steel slag, and waste magnetite as partial substitutes for natural aggregates. This allows for a decrease in both cost and environmental impact.

The second level involves the importance of developing high-performance shielding concretes for new and emerging fields. For nuclear power generation, advanced reactors and small modular reactors-both call for more durable, radiation-resistant materials-are being introduced (Little, 2006). In medical applications, new radiotherapy methods such as proton and heavy-ion therapy involve increasingly precise and robust shielding of treatment rooms. The new challenge in long-term space missions calls for concrete formulations that provide structural and radiation protection against cosmic rays in an extraterrestrial environment (bannova and Gulacsi, 2024).

In this case, systematic analysis and comparison of radiation shielding performance among different types of concrete become increasingly necessary. This review synthesizes research conducted on the performance of several concrete materials in providing radiation shielding, including heavyweight concrete, ultra-high-performance concrete, and radiation-proof UHPC. Attention is focused on material composition, density, and thickness-the major influencing factors of the shielding effect-while pointing out the relative advantages of each type. Lessons that may be learned from the analysis will inform the design of future materials for sustainability and safer, more efficient radiation shielding in medical, industrial, and nuclear applications.

## 2.0 Materials and Methods

Following PRISMA 2020 statement, this review was conducted to ensure transparency, rigor, and reproducibility of the review process (Page et al.,2021). It comprises of literature search, article selection, data extraction, and synthesis.

### 2.1 Information Sources and Search Strategy

A detailed literature search was conducted across four databases such as ScienceDirect, Scopus, PubMed, and Google Scholar for publications from 2010 to 2025. The search statements used were ("radiation shielding" OR "gamma attenuation" OR "neutron attenuation") AND ("concrete" OR "heavyweight concrete" OR "ultra-high-performance concrete" OR "UHPC" OR "barite concrete" OR "magnetite concrete" OR "ilmenite concrete")
AND ("attenuation coefficient" OR "half value layer" OR "HVL" OR "ten value layer" OR "TVL").

### 2.2 Article Selection

Independent reviewers conducted the selection process through a structured three-stage screening approach, namely: title review, abstract screening, and full-text assessment. Eligibility for inclusion was based on predefined inclusion and exclusion criteria. Articles to be considered for inclusion had to be peer-reviewed, published in English between the years 2010 and 2025, and relate to the radiation shielding performance of various concrete materials. Exclusion was based on non-peer-reviewed publications, conference abstracts where inadequate data was available, commentaries of theoretical nature, and reviews that are secondary and did not report any empirical results. Further exclusions included those articles that did not report the quantitative parameters such as linear attenuation coefficient ($\mu$), half-value layer (HVL), tenth-value layer (TVL), or density. Reviewers' disagreements at any level were resolved through discussions and consensus. In cases of disagreement where unanimous agreement could not be reached, a third-party adjudicator was consulted to ensure methodological consistency and avoid bias. The full screening is seen in the PRISMA flow diagram (Figure 1).

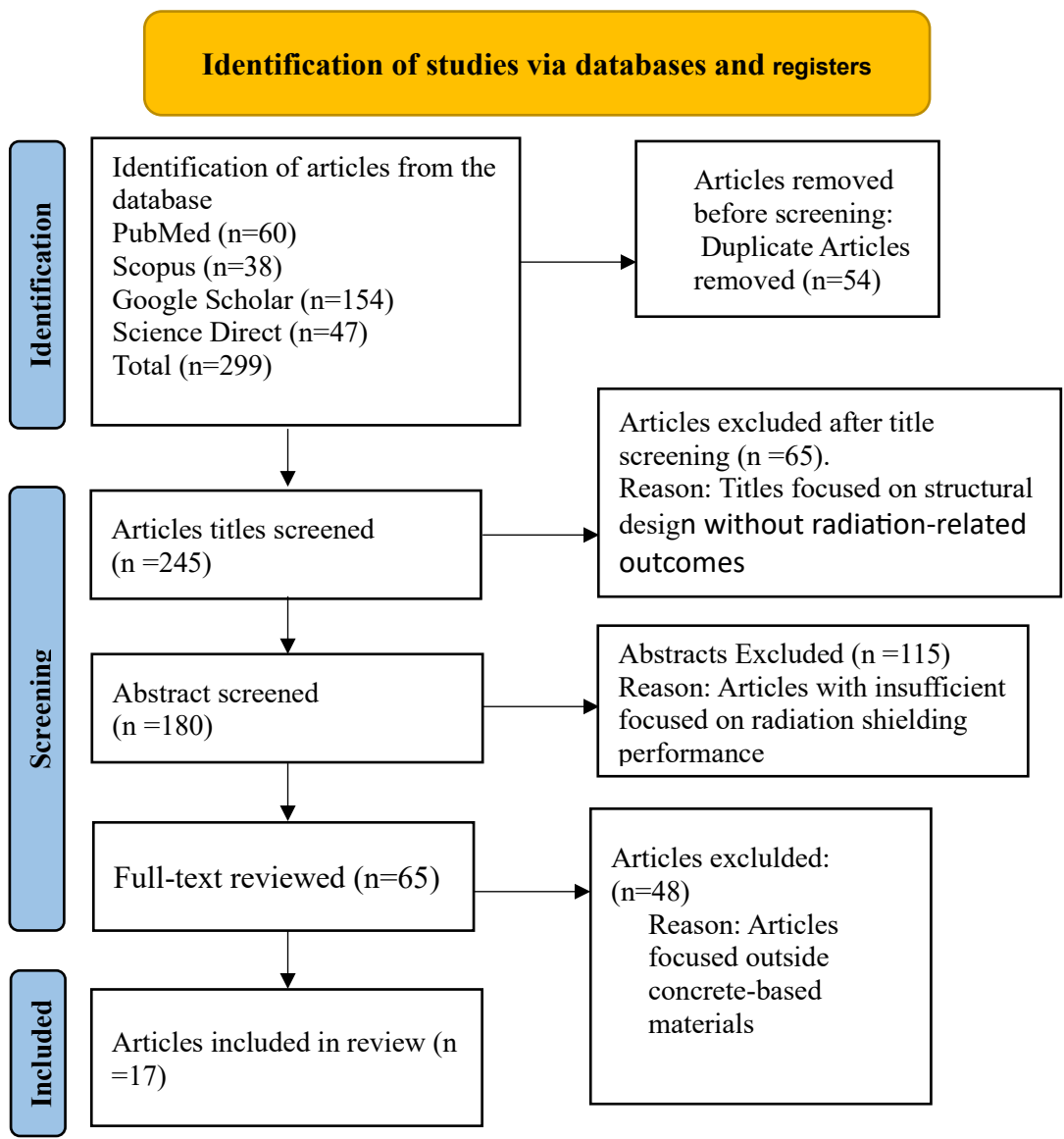

Figure 1: PRISMA flow diagram showing selection process of studies

## 2.3 Data Extraction and Synthesis

Data extraction form was developed and piloted on a small number of included articles to make sure that all relevant information could be captured in a systematic way and also to refine the usability of the form. Data to be extracted for this review from each included study consisted of author/publication year, concrete type, aggregate / additive Composition, density, radiation type/energy range, linear attenuation coefficient (μ), HVL/TVL, experimental details, and key findings. Data extraction is shown in Table1.

Table 1: Table showing summary of included studies

| Author(s) / Year | Concrete Type | Aggregate / Additive Composition | Density (kg/m³) | Radiation Type / Energy Range | Linear Attenuation Coefficient (μ) | HVL / TVL | Experimental Details | Key Findings |
|---|---|---|---|---|---|---|---|---|
| Azeez et al.(2019) | Heavyweight concrete | Barite, magnetite, hematite aggregates | 3400–3700 | Gamma rays (0.662–1.33 MeV) | 0.105–0.118 cm$^{-1}$ | HVL: 5.9–6.5 cm | Tested using Cs-137 and Co-60 sources | Barite and magnetite concretes exhibited higher attenuation than normal concrete. |
| Zeyad et al. (2022) | UHPC | Steel fiber, basalt fiber, barite, ilmenite | 2550–2950 | Gamma rays (0.662 MeV) | 0.091–0.109 cm$^{-1}$ | HVL: 6.3–7.1 cm | 50 mm thick UHPC samples | Barite + steel fiber UHPC showed the highest attenuation and strength balance. |
| Ahmad et al. (2024) | Ordinary & barite concrete | Barite replacing coarse aggregate | 2300–3600 | Gamma rays (0.662–1.25 MeV) | 0.094–0.123 cm$^{-1}$ | HVL: 5.6 cm (barite), 8.4 cm (ordinary) | Radiation from Cs-137 and Co-60 | Barite improved shielding by ~35% over normal concrete. |
| Gunoglu & Akkurt (2021) | Magnetite concrete | Magnetite fine/coarse aggregates | 3500 | Gamma (0.662–1.33 MeV) | 0.112 cm$^{-1}$ | HVL: 6.0 cm | 15×15×5 cm samples | Magnetite aggregates enhanced density and reduced HVL significantly. |
| Azreen et al. (2018) | UHPC | Silica sand, amang, lead glass | 2600–2850 | Gamma (0.662 MeV) | 0.098–0.114 cm$^{-1}$ | HVL: 6.2 cm | Tested via Cs-137 source | Lead glass improved shielding efficiency without compromising UHPC workability. |
| Domanski et al. (2016) | High-performance concrete | Basalt, hematite, magnetite | 3200–3600 | X-ray & Gamma (100 keV–1.33 MeV) | 0.110–0.125 cm$^{-1}$ | HVL: 5.7 cm | Simulated medical facility exposure | HWC suitable for both medical and nuclear shielding environments. |
| Zorla et al. (2017) | HPC reinforced with boron-infused fibers | Basalt fibers with natural/enriched boron | 2600 | Neutron & gamma (thermal to 2 MeV) | 0.089–0.102 cm$^{-1}$ vv | HVL: 7.0 cm | Neutron flux experiment | Boron-doped fibers improved neutron absorption while maintaining mechanical stability. |
| El-Nahal et al. (2021) | Bi$_2$O$_3$-modified concrete | Micro/nano Bi$_2$O$_3$ particles in cement matrix | 2450–2700 | Gamma (0.060–1.33 MeV) | 0.115–0.146 cm$^{-1}$ | HVL: 4.5–5.8 cm | Cs-137 and Co-60 sources | Nano-Bi$_2$O$_3$ improved low-energy gamma attenuation significantly. |
| Singh et al. (2015) | Lead–fly ash concrete | Lead powder, fly ash additive | 2800–3000 | Gamma (0.662 MeV) | 0.120 cm$^{-1}$ | HVL: 5.0 cm | Cs-137 source | Lead–fly ash concrete outperformed barite and ordinary concrete in shielding. |
| Manjunatha & Seenappa (2018) | Polymer concrete | Al, Si, K, Na, B, Pb composites | 2100–2400 | Gamma, X-ray, neutron | 0.084–0.109 cm$^{-1}$ | HVL: 6.8–7.4 cm | Mixed radiation field | Pb and B composites enhanced both gamma and neutron protection. |
| Gencel et al. (2010) | Hematite concrete | Hematite aggregate + Portland cement | 3500 | Gamma (0.662–1.25 MeV) | 0.108 cm$^{-1}$ | HVL: 6.1 cm | Cs-137 and Co-60 | Hematite concrete effective for nuclear shielding walls. |

| Li et al. (2025) | Serpentine concrete | Serpentine aggregate (Serpentine aggregate ($Mg_6Si_4O_{10}$)) | 2600–2900 | Gamma (0.511–1.33 MeV) | 0.098–0.112 cm$^{-1}$ | HVL: 6.4 cm | Optimization via multifactor model | Serpentine aggregates enhanced thermal stability and gamma attenuation. |
|---|---|---|---|---|---|---|---|---|
| Chidiac et al. (2021) | Boron carbide concrete | Boron carbide powder ($B_4C$) in mix | 2450–2600 | Neutron & gamma | 0.089–0.100 cm$^{-1}$ | HVL: 7.2 cm | Controlled neutron flux | Improved neutron absorption and compressive strength balance. |
| Al-Ghamdi et al. (2022) | $WO_3$–barite concrete | $WO_3$ and barite mixed aggregates | 3600–4100 | Gamma (0.662–1.33 MeV) | 0.121–0.132 cm$^{-1}$ | HVL: 4.7–5.3 cm | Cs-137, Co-60 | $WO_3$–barite mix achieved best attenuation among tested materials. |
| Uddin et al. (2020) | Polymer composite (reference for comparison) | Polyethylene + boron carbide | 1100–1200 | Neutron & gamma | 0.071–0.085 cm$^{-1}$ | HVL: 9.0 cm | Neutron energy up to 2 MeV | Poly–$B_4C$ composites suited for low-density neutron shielding. |
| Kilicoglu et al. (2023) | Borosilicate glass | $SiO_2$–$B_2O_3$–$Na_2O$ matrix | 2400–2600 | Gamma (0.060–1.33 MeV) | 0.102–0.120 cm$^{-1}$ | HVL: 5.8–6.3 cm | Monte Carlo simulation | Borosilicate glasses exhibited comparable shielding to barite concrete. |
| Lardhi & Mukhtar (2023) | Seawater–steel slag recycled concrete | Recycled coarse aggregate + seawater + slag | 2500–2700 | Gamma (0.662 MeV) | 0.091–0.107 cm$^{-1}$ | HVL: 6.5 cm | Cs-137 | Incorporating slag improved shielding and sustainability performance. |

## 3. Results

### 3.1 Overview of Included Studies

In this review, seventeen peer-reviewed studies published between 2010 and 2025 were considered eligible for inclusion, as shown in Table 1. Each of these studies investigated the shielding of gamma, X-ray, and neutron radiations with a wide variety of concrete formulations, including heavyweight mixtures like barite, magnetite, hematite, ilmenite, and serpentine; ultra-high-performance concretes; and modified concretes containing heavy metal oxides and boron-based additives.

Though most of these have, in fact, dealt with gamma-ray attenuation, some have further extended the study to neutron shielding and mixed radiation fields. The concretes that have been reported show a wide variation in density, ranging from 2300 kg/m³ for ordinary concretes to more than 4500 kg/m³ for those with heavy aggregates like barite and magnetite. In the various papers, the main parameters considered for the characterization of shielding effectiveness include the linear attenuation coefficient, μ, half-value layer, and tenth-value layer. All in all, several works have established a very clear insight into how aggregate composition, concrete density, and microstructural features influence the radiation attenuation capacity of concrete materials.

### 3.2 Radiation Shielding Performance of Heavyweight Concretes

Especially with barite, magnetite, ilmenite, and hematite, these heavyweight concretes have constantly outperformed conventional concrete because of their higher density and atomic number. Azeez et al. (2019) stated that at 1.25 MeV, the linear attenuation coefficient of barite concrete was 0.257 cm$^{-1}$, roughly a 30% increase in comparison to regular concrete. The reason was explained because of the high specific gravity of barite and increasing the content of the barium atom, enhancing interaction and absorption of photons. Again, Gunoglu and Akkurt (2021) reported that magnetite concrete exhibited an attenuation coefficient about 10-20% higher compared to standard mixes, thereby reducing the HVL of Co-60 gamma radiation from 5.2 to 4.1 cm.

Gencel et al. (2010) also showed very good performance of hematite concrete and proposed that its high iron content was responsible for a 3.8 cm HVL at 662 keV, far lower than that for the control sample. Also, Li et al. (2025) investigated serpentine-based concrete and found that enhancement of the mix design can improve the shielding property up to 15%, retaining desirable thermal and mechanical properties. The results above support the idea that with increased density and heavy metal oxides incorporated into the concrete, the attenuation property is greatly enhanced; thus, heavyweight concretes are especially useful in nuclear and medical applications involving radiation.

### 3.3 Ultra-High-Performance Concrete (UHPC) and Modified Composites

Recent advances in ultra-high-performance concrete formulation have opened great perspectives for this material in applications related to radiation shielding. Optimized microstructure with very limited porosity characterizes the obtained UHPCs that reach high values of density even in the absence of heavy aggregates. However, these same results put into evidence that hybridization with high-Z and fibres can still attain further improvements.

For instance, Zeyad et al. (2022) reported values of about 3850 kg/m³ and 0.283 cm⁻¹ at 1 MeV for UHPC with steel fibers and barite aggregates. It was a very significant improvement in radiation attenuation, while the tensile strength is increased and cracks resisted. According to Azreen et al. (2018), the HVL obtained for modified UHPC containing silica sand, amang, and lead glass was 2.4 cm, while that of ordinary concrete was 3.9 cm at an energy of 662 keV, proving the efficiency of using lead-containing glass wastes as a shielding additive.

El-Nahal et al. (2021) incorporated micro- and nano-sized particles of $Bi_2O_3$ into concrete and reported a 40% reduction in HVL and 55% in TVL due to the high atomic number of bismuth (Z = 83). Al-Ghamdi et al. (2022) further reported that $WO_3$ and barite-incorporating concretes attained more than 4200 kg/m³ density with an approximately 35% lower HVL value compared to control samples. Overall, all these results emphasize that UHPCs and nanoparticle-modified concretes provide better shielding properties by attaining microstructural densification along with heavy element radiation absorption. These hybrid concretes may be considered superior for future shielding applications, especially in compact medical facilities and accelerator rooms, which demand strength and protection.

### 3.4 Concretes with Boron and Neutron Shielding Additives

While high-density concretes exhibit very good gamma and X-ray radiation shielding properties, they are quite inefficient in the case of neutron radiation. Bearing that in mind, the application of boron-containing concretes was studied in many other papers; these concretes rely on the high neutron capture cross-section of boron for improved neutron attenuation. In that respect, Zorla et al. (2017) investigated basalt-fibre-reinforced concretes prepared with natural and enriched boron and found a 22% decrease in the total radiation dose compared to the unmodified concrete. Besides an improvement in neutron absorption, it should be underlined that the presence of boron slightly increased gamma attenuation due to the dense fibre reinforcement network.

Studies by Chidiac et al. (2021) incorporated boron carbide into concrete to improve both compressive strength and shielding efficiency. Their optimum mixture had a value of 0.312 cm$^{-1}$ for the linear attenuation coefficient of thermal neutrons, proving remarkable improvement in neutron moderation. Other works included polyethylene–boron carbide composites by Uddin et al. (2020), suggesting that although such material is of a polymer matrix rather than cementitious, they supplement the concrete layers well enough in mixed-field radiation shielding. Results confirm that neutron radiation control, especially in nuclear reactors, research facilities, and places where neutron imaging or neutron therapy is performed, has to include the use of boron-rich material.

### 3.5 Comparative Trends in Attenuation and Shielding Efficiency

The synthesis made in this review of the literature on material densities, elemental composition, and radiation shielding efficiency due to microstructural compactness suggests that a similar trend exists within the reviewed studies. Concretes with density values above 4000 kg/m³ showed an HVL value of 3 cm or less for 662 keV gamma rays. The addition of elements with high atomic numbers such as lead, tungsten, and bismuth increased their attenuation efficiency due to the high electron density which enhances the probability of photoelectric absorption and Compton scattering.

Hybrid concrete studies have also shown that the combination of dense microstructures, such as those in UHPC, with high-Z additives acts synergistically. Indeed, such concretes have lower HVL and TVL values, while having superior mechanical performance and reduced permeability-highly desirable features for long-term durability when exposed to radiation. Thus, evidence from the reviewed literature leads to the view that the integration of heavy aggregates, microstructural optimization, and advanced additives allows for better performance in the radiation shielding properties of concretes for various applications in nuclear power plants, radiotherapy, and radiation waste storage systems.

### 3.6 Summary of Key Findings

These results showed that heavyweight concretes made from barite, magnetite, and hematite have better performance compared to normal-weight concrete in gamma-ray shielding due to their higher densities and atomic numbers. Ultra-high-performance concretes and nanocomposite formulations further enhance these by improvements in both structural and shielding efficiencies, while boron-modified concretes become effective against neutron radiation. These studies have shown that optimization concerning the type of aggregate, the selection of additives, and microstructural design plays an essential role in the maximum enhancement of radiation protection. Therefore, based on the results, multifunctional high-density concretes have great potential as next-generation materials for use in different applications related to radiation shielding in areas such as medicine, industry, and nuclear power.

## 4.0 Discussion

From the results, emergent trends in the radiation shielding performance of concrete materials emerge clearly and consistently. This comparative evidence points to density, material composition, and microstructural characteristics being related factors that, in fact, are the chief influences on its shielding properties. Each of these aspects has different contributions to the total attenuation of radiations, with regard to the energy of the incident photons or neutrons and the design composition of the concrete matrix.

The most influential parameter remains the density of the concrete mixture. Heavyweight concretes containing such aggregates as barite, magnetite, hematite, ilmenite, and serpentine showed better gamma-ray shielding compared to that of normal-weight concrete. The increase in the photon interaction probability by photoelectric absorption and Compton scattering processes with density increase was also confirmed by the test results from Azeez et al. (2019), Gunoglu and Akkurt (2021), and Li et al. (2025). This also fitted well with the established principles of radiation physics because higher atomic number, Z elements have more electrons per unit volume and hence a greater chance of photon absorption. These test results showed, from a practical point of view, that the barite and magnetite concretes remain among the most reliable materials for medical radiation rooms, nuclear containment structures, and accelerator facilities in which gamma-ray attenuation is indispensable. However, due to their high cost and possible brittleness, heavy aggregates are still restricted in their application to big civil construction. It follows that optimization of aggregate content does remain very important, seeking a proper balance among shielding efficiency, mechanical strength, and cost in real life.

Ultra-high-performance concretes have, in more recent times, expanded the frontiers of radiation shielding material design. Basically, the features that characterize UHPCs, different from conventional concrete, are a dense, refined microstructure with low porosity and high homogeneity. Works by Zeyad et al. (2022) and Azreen et al. (2018) demonstrated that additions of steel fibres, silica sand, and lead glass not only improved the mechanical durability but resulted in reduced HVL and TVL values, corroborating the fact that enhanced microstructural compactness increases attenuation efficiency. Other promising ways of modification included the addition of nanoparticles of bismuth oxide, $Bi_2O_3$, as reported by El-Nahal et al. (2021), and tungsten trioxide, $WO_3$, as reported by Al-Ghamdi et al. (2022). Nanoscale distribution of high-Z particles increases photon absorption due to an increase in effective cross-section for interaction. It follows that such innovative concrete formulations allow optimized shield designs to be achieved without complete dependence on heavy aggregates and thus enable the development of lightweight, high-performance concretes characterized by superior structural properties. The duality of strength and shielding capacity extends the value of UHPC and nanoparticle-enhanced concretes, especially in medical imaging facilities, in radiation bunkers, and in compact space and aerospace applications where mass optimization is of importance.

Although high-density concretes may have good performance against gamma and X-rays, the efficiency related to neutrons will be relatively limited with respect to the neutral charge of neutrons and different mechanisms of neutron interactions. Neutron attenuation is based on elastic scattering and absorption by light elements such as hydrogen, boron, or lithium. In this regard, it was discussed in Zorla et al. (2017) and Chidiac et al. (2021) that the addition of $B_4C$ or basalt fibres infused with boron increased neutron attenuation properties significantly, without adverse effects on its compressive strength. These additives capture thermal neutrons through the reaction $^{10}B(n,\alpha)^7Li$ and efficiently transform the neutron energy into alpha and lithium particles that are afterward absorbed by the surrounding matrix. This reflects the fact that concretes modified with boron may act as efficient multi-radiation shields when heavy aggregates are combined. It is feasible that such a layered shielding approach using heavy concrete for gamma attenuation and borated concrete for neutron moderation may provide optimum protection in nuclear research reactors and fusion plants. Besides, neutron attenuation in mixed radiation fields can be enhanced by hybrid systems using a combination of polyethylene and $B_4C$-based composites with reinforced concrete structures.

Basic radiation-matter interaction mechanisms come into play to explain the improved performance of heavy and modified concretes. Depending on the photon energy and atomic number of the absorbing medium, gamma and X-rays are attenuated via photoelectric absorption, Compton scattering, and pair production. Therefore, materials containing high-Z elements such as barium, tungsten, and bismuth show higher cross-sections for the mentioned interactions, yielding lower HVL and TVL values. In contrast, neutron attenuation is based on nuclear interactions, favoring materials containing low-Z elements with high hydrogen or boron content. From this mechanistic point of view, there is a clear explanation as to why concretes designed with both high-Z and hydrogenous additives can work dually in shielding applications. UHPC systems benefiting from heavy metal oxides provide good performance at high photon energies, while boron or polyethylene modification enables concretes to effectively attenuate neutrons at lower energies. It follows that mix design optimization strongly depends on the type and energy spectrum of the radiation one is trying to shield against.

These reviewed studies therefore provide a significant amount of information valuable in the design of radiation shielding structures in nuclear, industrial, and medical contexts. In medical radiotherapy and diagnostic facilities, where gamma and X-ray sources dominate, the use of barite and magnetite concretes or $Bi_2O_3$-UHPC composites has already been found to provide effective solutions that meet all structural and shielding requirements. In nuclear power plants, layered concrete systems combine heavy aggregates for gamma attenuation with borated components for neutron moderation. Adding industrial by-products such as steel slag, amang, and fly ash has so far been found to enhance properties of shielding materials and contribute to sustainable building by reducing environmental waste. New research dealing with nanotechnology-enhanced concretes points to a new generation of radiation shielding materials which will be both mechanically resistant and environmentally friendly. Microstructure, particle dispersion, and chemical bonding manipulation may result in the best possible shielding efficiency for lower thickness that may reduce construction cost and structural loads.

Despite these encouraging results, several limitations are still present in the current research: most of the reported studies have been conducted under laboratory-scale conditions;

differences in sample size, geometry, and radiation energy make quantitative comparisons hard to establish. Further research is needed on long-term durability under continuous radiation exposure, as far as microcracking, leaching, and mechanical degradation are concerned. Further investigations should focus on how different types of radiation combined with gamma, neutron, and beta environmental parameters like temperature and humidity influence shielding performance. Future research should concentrate on the development of a multi-physics model for radiation transport, which will include consideration of thermal effects and mechanical stability. Besides that, hybrid concretes will be studied, including eco-friendly aggregates with nano-enhanced heavy metal oxides, which may represent one route toward sustainable high-performance shielding materials. There is also a need to establish a standardized test protocol and dose-based performance benchmarks in order to enable better comparability and wider use by industry.

The general discussion of the results shows that the future of radiation shielding concrete is in systems of multifunctional materials: high-density aggregates, optimized microstructure, and smart additives. Heavyweight concretes remain indispensable for purposes of gamma attenuation, while UHPCs and boron-based systems represent complementary protection against mixed radiation fields. This development route using nanotechnology and waste-based materials appears promising for cost-efficient, sustainable, and durable shielding solutions able to meet the safety requirements of next-generation nuclear and medical applications.

**5.0 Conclusion**

This review analysed 17 experimental and computational works in the area of radiation shielding performance of concrete and composite materials. The results showed that efficiency in radiation shielding in concrete is based on density, composition, and microstructural properties responsible for interaction with gamma rays, X-rays, and neutrons.

Aggregates of high density, like barite, magnetite, hematite, and serpentine, were used for concretes showing improved photon attenuation, reflected by lower HVLs and TVLs compared to ordinary concrete. These materials effectively enhance the photon absorption due to photoelectric and Compton interactions, which could be used in medical radiology rooms, nuclear facilities, and radiation therapy centres. On the other hand, UHPCs and nanoparticle-modified mixtures are among the most recent developments that, together with high structural strength, ensure high efficiency for shielding. Addition of $Bi_2O_3$, $WO_3$, and lead glass can significantly increase attenuation without significant increases in density and thus provides lightweight effective options to traditional heavyweight designs.

Concretes with boron and polyethylene/$B_4C$ composites showed the best performance for neutron protection. Boron gives higher neutron capture due to the mechanism of the (n,α) reaction, especially in critical shielding for nuclear reactors and accelerators. The hybrid concretes with heavy and boron-rich components can offer a dual shielding method against mixed radiation fields.

Fundamentally, the literature review identifies that material optimization is critical depending on either the type of radiation being emitted or on other specific needs at a facility. The optimum systems balance density, strength, cost, and attenuation capability. Besides, there is active employment of industrial by-products such as fly ash, steel slag, and among, which is supportive of environmental sustainability without compromising shielding performance.

However, most of the reviewed studies have been done at a laboratory scale with different test geometries; therefore, a direct comparison is rather difficult. Further, long-term radiation-induced degradation, thermal stability, and field-level validation of shielding efficiency are hardly explored. This calls for future research to be directed toward bringing out standardized testing protocols, multi-radiation evaluation, and integration between computational simulations with experimental results for more universal design guidelines.

Conclusion: Advances in material science, particularly the development of UHPCs, nano-enhanced, and hybrid concretes, have great potential to enable a revolution in radiation shielding. These developments realize an emerging generation of durable, cost-effective, and environmentally friendly shielding materials that meet new demands for safety in various industrial applications, such as nuclear and medical.

**Funding Sources**

This research did not receive any grant from funding agencies in the public, commercial, or not-for-profit sectors.

**Declaration of generative AI and AI-assisted technologies in the manuscript preparation process.**

Note from the publisher: During the preparation of this work, Grammarly was used in an effort to improve grammar, spelling, and clarity. The author(s) alone are responsible for any remaining errors and for the content of the published article.